# EDITORIAL: STATISTICS AND "THE LOST TOMB OF JESUS"

By Stephen E. Fienberg

*Carnegie Mellon University*

What makes a problem suitable for statistical analysis? Are historical and religious questions addressable using statistical calculations? Such issues have long been debated in the statistical community and statisticians and others have used historical information and texts to analyze such questions as the economics of slavery, the authorship of the Federalist Papers and the question of the existence of God. But what about historical and religious attributions associated with information gathered from archeological finds?

In 1980, a construction crew working in the Jerusalem neighborhood of East Talpiot stumbled upon a crypt. Archaeologists from the Israel Antiquities Authority came to the scene and found 10 limestone burial boxes, known as ossuaries, in the crypt. Six of these had inscriptions. The remains found in the ossuaries were reburied, as required by Jewish religious tradition, and the ossuaries were catalogued and stored in a warehouse. The inscriptions on the ossuaries were catalogued and published by Rahmani (1994) and by Kloner (1996) but there reports did not receive widespread public attention.

Fast forward to March 2007, when a television "docudrama" aired on *The Discovery Channel* entitled "The Lost Tomb of Jesus"[1] touched off a public and religious controversy—one only need think about the title to see why there might be a controversy! The program, and a simultaneously published book [Jacobovici and Pellegrino (2007)], described the "rediscovery" of the East Talpiot archeological find and they presented interpretations of the ossuary inscriptions from a number of perspectives. Among these was a statistical calculation *attributed to* the statistician Andrey Feuerverger: "that the odds that all six names would appear together in one tomb are 1 in 600, calculated conservatively—or possibly even as much as one in one million."

At about this time, Feuerverger submitted a paper to *The Annals of Applied Statistics* (AOAS) for review, but its contents remained confidential and only a rough outline of the details of his calculations was publicly available [Mims (2007)]. Commentary regarding Feuerverger's statistical calculation quickly appeared on the web. Was it really a Bayesian calculation? On what assumptions were the statistical arguments based? Most criticism focused not directly on the actual statistical arguments but on how they were portrayed by the documentary's producers and interpreted by others. And the controversy over the broader interpretation and claims regarding the origin of the East Talpiot tomb raged on.

[1] "The Lost Tomb of Jesus." *Discovery Channel*, March 4, 2007. http://dsc.discovery.com/convergence/tomb/tomb.html.

In July 2007 at the Joint Statistical Meetings in Salt Lake City, Feuerverger gave the first public airing of the details of his work and three discussants presented alternative perspectives. The paper itself underwent an extensive review process and a substantially revised version appears in this issue of AOAS [Feuerverger (2008)]. It includes photographs, detailed discussion of possible data on names from ancient sources, the assumptions upon which the analysis was based, and a novel $p$-value calculation. The paper is accompanied by a series of detailed discussions and critiques, several of which reframe the statistical problem from a Bayesian perspective.

The AOAS editors encourage our readers to judge for themselves the persuasiveness of the assumptions, the data, and the calculations performed by Feuerverger, especially in light of the criticisms voiced in the extended discussion that follows his paper, and his response. Interested readers may then wish to explore the extensive nonstatistical discussion of the East Talpiot available in print and on the web.

DEPARTMENT OF STATISTICS AND MACHINE LEARNING
CARNEGIE MELLON UNIVERSITY
PITTSBURGH, PENNSYLVANIA 15213
USA
E-MAIL: fienberg@stat.cmu.edu